\documentclass[aps,prd,nofootinbib,amsmath,amssymb,superscriptaddress,twocolumn,10pt]{revtex4}

\pdfoutput=1

\usepackage{graphicx}
\usepackage{dcolumn}
\usepackage{bm}
\usepackage{amssymb}
\usepackage{latexsym}
\usepackage{booktabs}
\usepackage{amsmath}
\usepackage{multirow}
\usepackage[colorlinks=true, linkcolor=red, citecolor=blue]{hyperref}

\usepackage[usenames,dvipsnames]{xcolor}

\newcommand{\be}{\begin{equation}}
\newcommand{\ee}{\end{equation}}
\newcommand{\bq}{\begin{eqnarray}}
\newcommand{\eq}{\end{eqnarray}}

\begin{document}

\title{Cosmological search for sterile neutrinos after Planck 2018}

\author{Lu Feng}
\affiliation{College of Physical Science and Technology, Shenyang Normal University, Shenyang 110034, China}
\affiliation{Department of Physics, College of Sciences, Northeastern University, Shenyang 110819, China}
\author{Rui-Yun Guo}
\affiliation{School of Sciences, Xi'an Technological University, Xi'an 710021, China}
\author{Jing-Fei Zhang}
\affiliation{Department of Physics, College of Sciences, Northeastern University, Shenyang 110819, China}
\author{Xin Zhang\footnote{Corresponding author}}
\email{zhangxin@mail.neu.edu.cn}
\affiliation{Department of Physics, College of Sciences, Northeastern University, Shenyang 110819, China}
\affiliation{Frontiers Science Center for Industrial Intelligence and Systems Optimization, Northeastern University, Shenyang 110819, China}
\affiliation{Key Laboratory of Data Analytics and Optimization for Smart Industry (Northeastern University), Ministry of Education, China}

\begin{abstract}
Sterile neutrinos can affect the evolution of the universe, and thus using the cosmological observations can search for sterile neutrinos. In this work, we use the cosmic microwave background (CMB) anisotropy data from the Planck 2018 release, combined with the latest baryon acoustic oscillation (BAO), type Ia supernova (SN), and Hubble constant ($H_0$) data, to constrain the cosmological models with considering sterile neutrinos. In order to test the influences of the properties of dark energy on the {results} of searching for sterile neutrinos, in addition to the $\Lambda$ cold dark matter ($\Lambda$CDM) model, we also consider the $w$CDM model and the holographic dark energy (HDE) model. We find that the existence of sterile neutrinos {is not preferred} when the $H_0$ local measurement is not included in the data combination. When the $H_0$ measurement is included in the joint constraints, it is found that $\Delta N_{\rm eff}>0$ is {favored} at about 2.7$\sigma$ level for the $\Lambda$CDM model and at about 1--1.7$\sigma$ level for the $w$CDM model. However, $m_{\nu,{\rm{sterile}}}^{\rm{eff}}$ still cannot be well constrained and only upper limits can be given. In addition, we find that the HDE model is definitely ruled out by the current data. We also discuss the issue of the Hubble tension, and we conclude that involving sterile neutrinos in the cosmological models cannot truly resolve the Hubble tension.

\end{abstract}
\maketitle

\section{Introduction}
\label{sec1}
At present, the possible existence of sterile neutrinos is one of the most debated topics in neutrino physics. {The recent experiments and the historic anomalies}~\cite{Aguilar:2001ty,Acero:2007su,Mention:2011rk,Aguilar-Arevalo:2012fmn,Giunti:2012tn,Gariazzo:2013gua,Hayes:2013wra,
Hannestad:2012ky,Giunti:2019aiy,Gariazzo:2019gyi,Diaz:2019fwt,Boser:2019rta,Hagstotz:2020ukm,Dasgupta:2021ies} seem to point towards the existence of light massive sterile neutrinos with the mass around the eV scale, but some other experiments, {such as the result of neutrino oscillation experiment by the Daya Bay and MINOS collaborations~\cite{TheIceCube:2016oqi} and the result of cosmic ray experiment by the IceCube collaboration~\cite{Adamson:2016jku}}, did not detect such a signal, which casts doubt on this hypothesis. Since the sterile neutrinos have some effects on the evolution of the universe, cosmological observations can provide independent way to search for sterile neutrinos.

However, using the cosmological observations to search for sterile neutrinos depends on cosmological models. In order to fit the cosmological data, one needs to assume a specific cosmological model with some cosmological parameters, and these cosmological parameters could be simultaneously determined in the sense of statistics in the cosmological fit. Therefore, the cosmological searches of sterile neutrinos not only depend on cosmological observations, but also depend on cosmological models. In particular, some recent studies \cite{Feng:2017nss,Zhao:2017urm,Feng:2017mfs,Feng:2017usu,Feng:2019jqa} have revealed that the properties of dark energy can significantly impact on the cosmological {fit results} of sterile neutrinos.

On the other hand, currently, one of the most important puzzles in cosmology is the Hubble tension \cite{Verde:2019ivm}. It was found that a significant tension exists between the observations of the early and late universe. One way of relieving the Hubble tension is to consider a dynamical dark energy in a cosmological model \cite{Li:2013dha}. It has been found that the equation of state (EoS) of dark energy is in anti-correlation with the Hubble constant in the cosmological fits using the observation of cosmic microwave background (CMB) anisotropies. However, due to the observations becoming increasingly precise, the EoS of dark energy has been constrained tightly and thus cannot provide enough room for accommodating a rather high value of the Hubble constant \cite{Guo:2018ans}. Nevertheless, it is also well known that the existence of sterile neutrinos can also help relieve the Hubble tension, as the parameter $N_{\rm eff}$ of sterile neutrinos is in positive correlation with the Hubble constant. In this circumstance, an obvious further way is to simultaneously consider a dynamical dark energy and sterile neutrinos in a cosmological model \cite{Zhao:2017urm}. Although such a consideration indeed can effectively relieve the Hubble tension, a deep analysis shows that the current cosmological observations might not favor such models when using {the Akaike (or Bayesian) information criterion} to assess the fits \cite{Guo:2018ans}. Anyway, on one hand, the Hubble constant measurement is useful in searching for sterile neutrinos in cosmology, and on the other hand, the consideration of sterile neutrinos in a cosmological model is also helpful in solving the problem of the Hubble tension. See e.g. Refs.~\cite{Jacques:2013xr,Mirizzi:2013kva,Wyman:2013lza,Archidiacono:2014apa,Bergstrom:2014fqa,Leistedt:2014sia,Gariazzo:2015rra,Abazajian:2017tcc,
Choudhury:2018sbz,Hamann:2012fe,Wang:2012vh,Giusarma:2013pmn,Zheng:2014dka,Zhang:2014nta,Lesgourgues:2014zoa,Costanzi:2014tna,Zhang:2014ifa,
Qian:2015waa,Patterson:2015xja,Allison:2015qca,Geng:2015haa,Zhang:2015uhk,Huang:2015wrx,Hada:2016dje,Wang:2016tsz,Boehringer:2016bzy,Zhao:2016ecj,
Vagnozzi:2017ovm,Guo:2017hea,Yang:2017amu,Zhang:2017rbg,Chen:2017ayg,Koksbang:2017rux,Xu:2016ddc,Li:2017iur,Wang:2017htc,Zhao:2017jma,Boyle:2017lzt,
Vagnozzi:2018jhn,Wang:2018lun,Guo:2018gyo,Boyle:2018rva,Upadhye:2017hdl,Feng:2019mym,Li:2020gtk,Zhang:2020mox,Vagnozzi:2019utt,Amiri:2021kpp,
DiValentino:2021izs,Freedman:2021ahq,Schoneberg:2021qvd} for related studies.

Recently, the latest CMB data, i.e., the Planck 2018 angular power spectra data, have been released by the Planck collaboration, and some other important updated observational data, {including the baryon acoustic oscillation (BAO) data, the type Ia supernovae (SN) data, and the Hubble constant $H_0$ data,} have also been released. Thus, it is necessary to make a new analysis for the issue of searching for sterile neutrinos in cosmology using the cosmological observations.

In this work, we will use the latest CMB, BAO, SN, and $H_0$ data to search for sterile neutrinos. Since the influence of dark energy is important in this issue, we will not only assume a $\Lambda$ cold dark matter ($\Lambda$CDM) model, but also consider dynamical dark energy models in the cosmological fits. To be simple as far as possible, we only consider the simplest dynamical dark energy models in this work. Therefore, we only consider the $w$CDM model and the holographic dark energy (HDE) model in our analysis. These two models have only one extra parameter compared to $\Lambda$CDM. For the $w$CDM model, the EoS of dark energy $w$ is a constant. For the HDE model, the energy density of dark energy is given by $\rho_{\rm{de}}=3c^2 M^2_{\rm{pl}}R^{-2}_{\rm{eh}}$, where $c$ is a dimensionless parameter which plays an important role in determining the evolution of dark energy in the HDE model and $M_{\rm{pl}}$ is the reduced Planck mass. {$R_{\rm{eh}}$ is the future event horizon, defined as $R_{\rm{eh}}=a(t)\int_{t}^{\infty}\frac{dt'}{a(t')}=a\int_{a}^{\infty}\frac{da'}{Ha'^2}$, where $a(t)$ is the scale factor of our universe and $H=\dot{a}/a$ is the Hubble parameter, with the dot denoting the derivative with respect to the cosmic time $t$}. In the HDE model \cite{Li:2004rb}, the evolution of EoS is given by $w(a)=-1/3-(2/3c)\sqrt{\Omega_{\rm de}(a)}$. In this case, the only extra parameter relative to $\Lambda$CDM is the parameter $c$. For more details of the HDE model, see e.g., Refs.~\cite{Huang:2004wt,Huang:2004mx,Wang:2004nqa,Zhang:2005hs,Nojiri:2005pu,Chang:2005ph,Zhang:2007sh,Ma:2007av,Li:2008zq,Ma:2007pd,Li:2009bn,Zhang:2009xj,delCampo:2011jp,Li:2013dha,Cui:2015oda,Landim:2015hqa,Xu:2016grp}.
Although the two dynamical dark energy models are simple, they are rather representative. In the $w$CDM model, the dark energy is either quintessence type ($w>-1$) or phantom type ($w<-1$). While in the HDE model, the EoS of dark energy is dynamically evolutionary, and the dark energy can be quintessence type ($c>1$) with $w$ always larger than $-1$ or quintom type ($c<1$) with $w$ evolving from $w>-1$ to $w<-1$.

In this work, we constrain the $\Lambda$CDM, $w$CDM, and HDE models where sterile neutrinos are considered using the latest cosmological observations, and we discuss the issues of {cosmological searches} of sterile neutrinos, impacts of properties of dark energy, model comparison, and Hubble tension, based on the {cosmological fit results}.

\section{Data and method}
\label{sec2}

\subsection{Data}

In this paper, for the observational data, we consider the following data sets.

{\it The CMB data}: We use the CMB likelihood including the TT, TE, EE spectra at $l\geq 30$, the low-$l$ temperature commander likelihood, and the low-$l$ SimAll EE likelihood from the Planck 2018 data release~\cite{Aghanim:2018eyx}.

{\it The BAO data}: We use the measurements from the six-degree-field galaxy survey sample ($z_{\rm eff}=0.106$)~\cite{Beutler:2011hx}, the Sloan Digital Sky Survey Main Galaxy Sample ($z_{\rm eff}=0.15$)~\cite{Ross:2014qpa}, and the Baryon Oscillation Spectroscopic Survey Data Release 12 ($z_{\rm eff}=$ 0.38, 0.51, and 0.61)~\cite{Alam:2016hwk}.

{\it The SN data}: We use the latest Pantheon sample, which is comprised of 1048 data points~\cite{Scolnic:2017caz}.

{\it The $H_0$ measurement}: We use the local measurement result of $H_0=74.03{\pm1.42}~{\rm km}~{\rm s}^{-1}~{\rm Mpc}^{-1}$ from the cepheid-supernova distance ladder, reported in Ref.~\cite{Riess:2019cxk}.

In what follows, we will use these observational data to place constraints on the $\Lambda$CDM, $w$CDM, and HDE models with and without sterile neutrinos. We will use two data combinations, i.e., CMB+BAO+SN (abbreviated as CBS) and CMB+BAO+SN+$H_0$ (abbreviated as CBSH), to constrain the cosmological parameters. These usages enable us to conveniently compare with {the cosmological fit results of the neutrinos mass} obtained in previous works, e.g., Refs.~\cite{Zhang:2020mox,Li:2020gtk}.

\subsection{Method}

In a cosmological model without considering sterile neutrinos, the free parameter vector is ${\bf P}=\{\Omega_bh^2,~\Omega_ch^2,~100\theta_{\rm MC},~\tau,~\ln (10^{10}A_s),~n_s,~w~({\rm or}~c)\},$
where $\Omega_bh^2$ and $\Omega_ch^2$ represent the physical baryon density and the physical cold dark matter density, respectively, $\theta_{\rm MC}$ is the ratio (multiplied by 100) between the sound horizon $r_s$ and angular diameter distance $D_{\rm A}$ at decoupling, $\tau$ is the optical depth to the reionization, $A_s$ is the amplitude of the power spectrum of primordial curvature perturbations, $n_s$ is the power-law spectral index, $w$ is the EoS parameter of dark energy for the $w$CDM model, and $c$ is the dimensionless phenomenological parameter for determining the evolution of dark energy in the HDE model. Thus, there are six independent parameters in total for the $\Lambda$CDM model and seven independent parameters in total for the $w$CDM model and the HDE model. The total mass of active neutrinos $\sum m_\nu$ is fixed at 0.06eV.

In this work, we consider the both cases of massless and massive sterile neutrinos. When the case of massless neutrinos (as the dark radiation) is considered in the cosmological models, one extra free parameter, the effective number of relativistic species $N_{\rm eff}$ should be involved in the calculation. When massless sterile neutrinos are considered in the $\Lambda$CDM model, the $w$CDM model, and the HDE model, these cases are called the $\Lambda$CDM+$N_{\rm eff}$ model, the $w$CDM+$N_{\rm eff}$ model, and the HDE+$N_{\rm eff}$ model, respectively. Thus, the $\Lambda$CDM+$N_{\rm eff}$ model has seven independent parameters, and the $w$CDM+$N_{\rm eff}$ model and the HDE+$N_{\rm eff}$ model have eight independent parameters.

When the sterile neutrinos are considered to be massive, two additional parameters, $N_{\rm eff}$ and the effective sterile neutrino mass $m_{\nu,{\rm sterile}}^{\rm eff}$, need to be added in the cosmological models. Correspondingly, the models considered in this paper are called the $\Lambda$CDM+$N_{\rm eff}$+$m_{\nu,{\rm sterile}}^{\rm eff}$ model, the $w$CDM+$N_{\rm eff}$+$m_{\nu,{\rm sterile}}^{\rm eff}$ model, and the HDE+$N_{\rm eff}$+$m_{\nu,{\rm sterile}}^{\rm eff}$ model, respectively. The $\Lambda$CDM+$N_{\rm eff}$+$m_{\nu,{\rm sterile}}^{\rm eff}$ model has eight independent parameters, and the $w$CDM+$N_{\rm eff}$+$m_{\nu,{\rm sterile}}^{\rm eff}$ and HDE+$N_{\rm eff}$+$m_{\nu,{\rm sterile}}^{\rm eff}$ models have nine independent parameters.

Note here that in the cases of considering sterile neutrinos the prior of $N_{\rm eff}>3.044$ should be set in the calculations.

We use the {\tt CosmoMC} package \cite{Lewis:2002ah} to infer the posterior probability distributions of the sterile neutrino parameters and other cosmological parameters.


The total $\chi^2$ of the two data combinations can be written as $\chi^2=\chi^2_{\rm CMB}+\chi^2_{\rm BAO}+\chi^2_{\rm SN}$ and $\chi^2=\chi^2_{\rm CMB}+\chi^2_{\rm BAO}+\chi^2_{\rm SN}$+$\chi^2_{H_0}$. In general, the $\chi^2$ comparison is simplest analysis method for comparing different models with the same parameter number. When the comparison is made for models with different numbers of free parameters,
{a model with more parameters tends to give a better fit to the same data ($\chi^2_{\rm min}$ tends to be smaller), and thus}
the simple $\chi^2$ comparison is obviously unfair. Therefore, in this work we use the Akaike information criterion (AIC) as an evaluation tool to compare different cosmological models. A model with a smaller value of AIC is believed to be more favored by data.
The AIC is defined as $\rm AIC=\chi^2_{\rm min}$+2$k$, where $k$ is the number of parameters. Actually, we only care about the relative values of AIC between different models, and thus we use $\Delta\rm AIC=\Delta\chi^2_{\rm min}$+$2\Delta k$ to compare models. Here we take the $\Lambda$CDM model as a reference model for calculating the $\Delta\rm AIC$ values for other models.

\section{Results and discussion}\label{sec3}

\begin{figure}[!htp]
\includegraphics[scale=0.7]{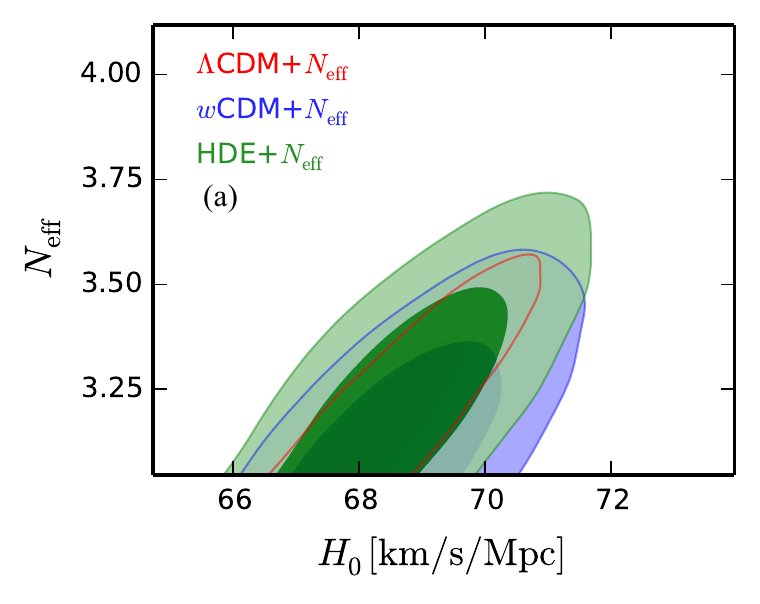}
\includegraphics[scale=0.7]{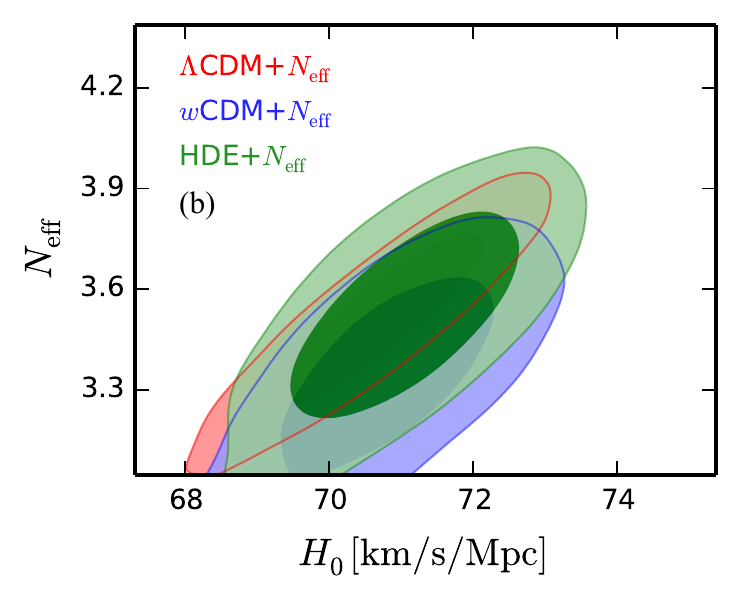}
\centering
 \caption{\label{figneff} Two-dimensional marginalized posterior contours (68.3\% and 95.4\% CL) in the $H_0$--$N_{\rm eff}$ plane for the $\Lambda$CDM+$N_{\rm eff}$ model, the $w$CDM+$N_{\rm eff}$ model, and the HDE+$N_{\rm eff}$ model by using the CBS (a) and CBSH (b) data combinations. }
\end{figure}

\begin{figure}[!htp]
\includegraphics[scale=0.5]{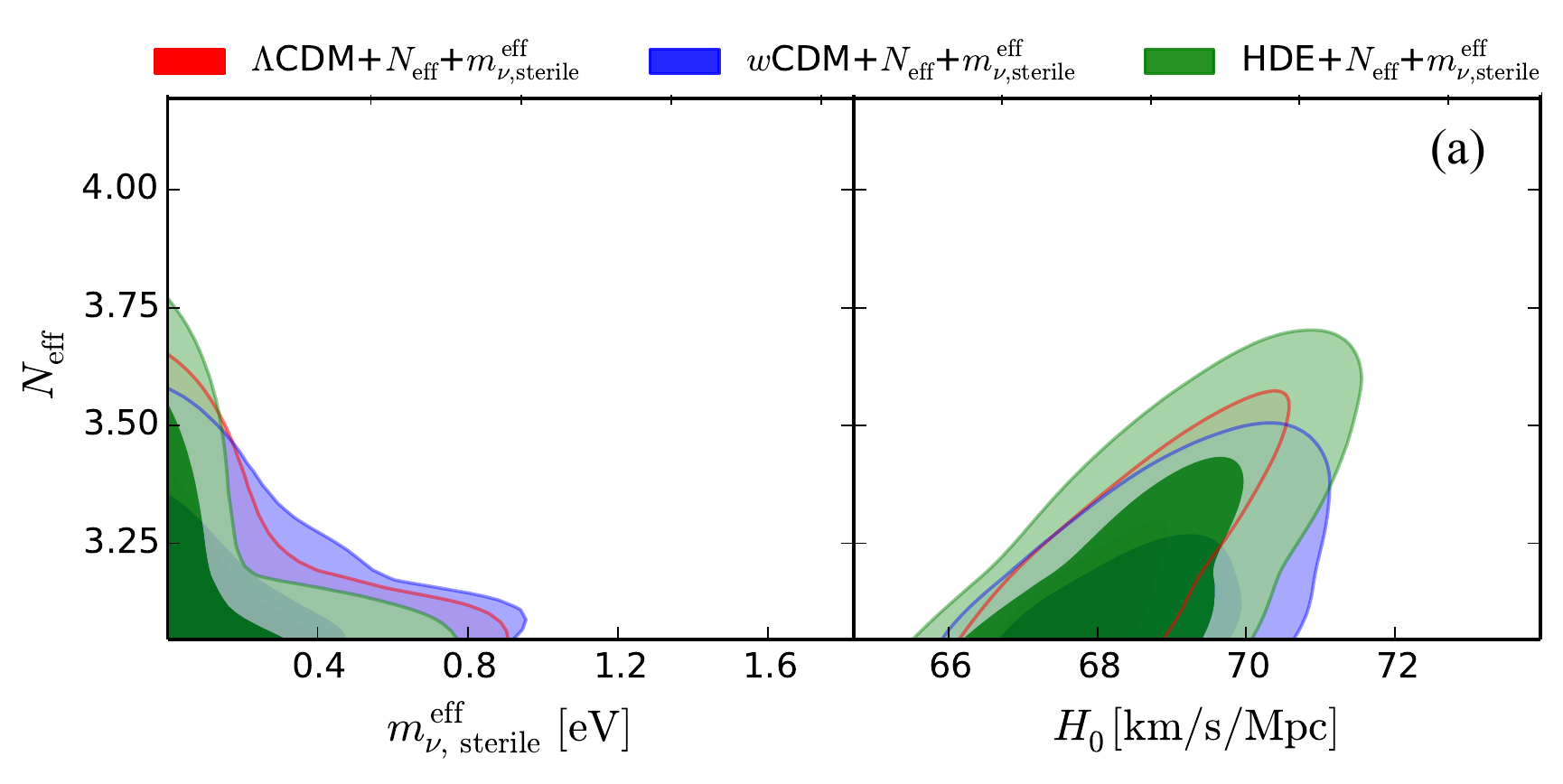}
\includegraphics[scale=0.5]{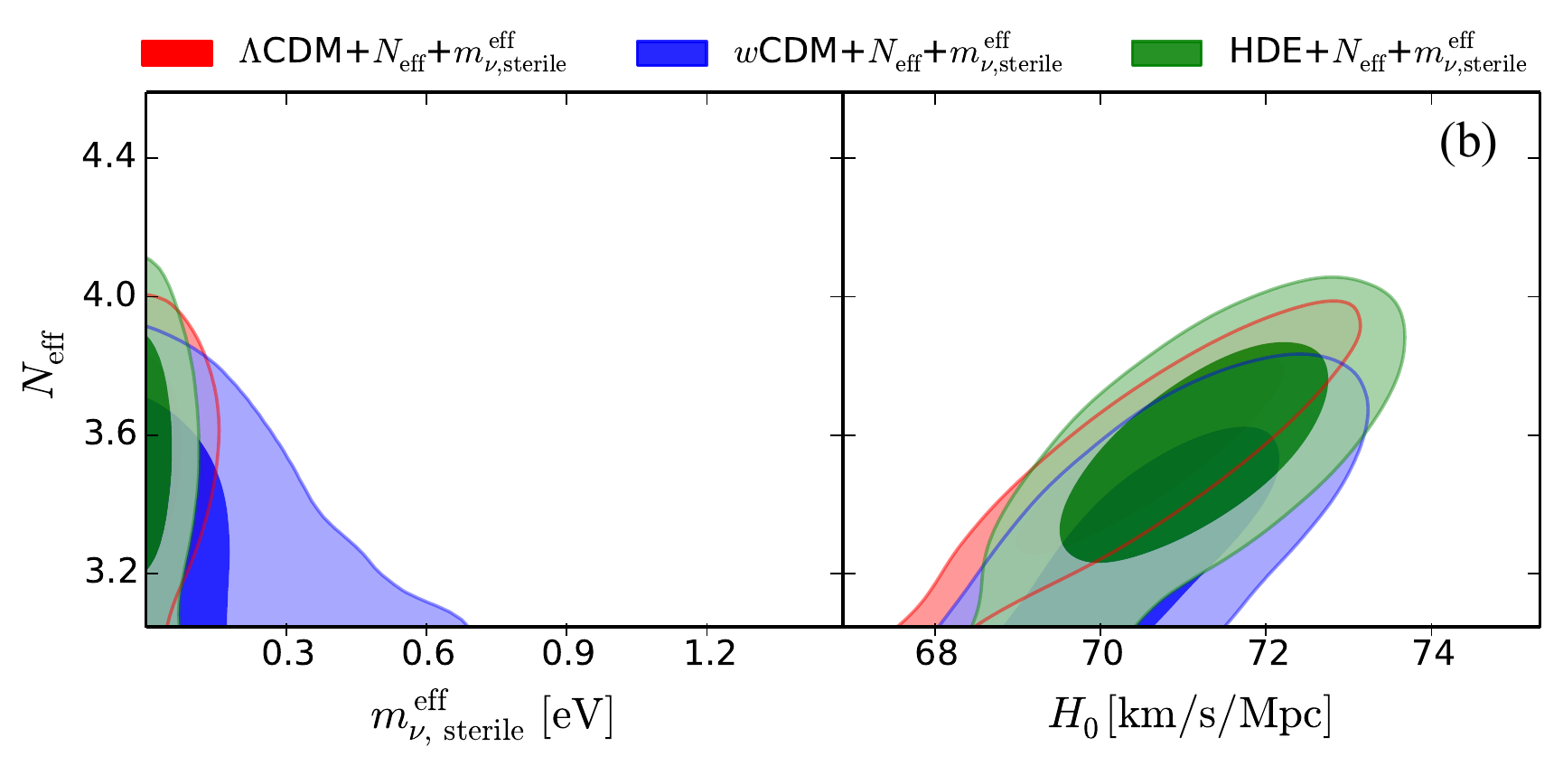}
\centering
 \caption{\label{figms} Two-dimensional marginalized posterior contours (68.3\% and 95.4\% CL) in the $m_{\nu,{\rm{sterile}}}^{\rm{eff}}$--$N_{\rm eff}$ and $H_0$--$N_{\rm eff}$ planes for the $\Lambda$CDM+$N_{\rm eff}$+$m_{\nu,{\rm{sterile}}}^{\rm{eff}}$ model, the $w$CDM+$N_{\rm eff}$+$m_{\nu,{\rm{sterile}}}^{\rm{eff}}$ model, and the HDE+$N_{\rm eff}$+$m_{\nu,{\rm{sterile}}}^{\rm{eff}}$ model by using the CBS (a) and CBSH (b) data combinations. }
\end{figure}

In this section, we report the fitting results of the cosmological models and discuss the implications of these results in the searches for sterile neutrinos using the latest observational data. The fitting results are given in Tables~\ref{tabcbs} and \ref{tabcbsh} as well as Figs.~\ref{figneff} and \ref{figms}.

\begin{table*}\small
\setlength\tabcolsep{0.2pt}
\renewcommand{\arraystretch}{1.2}
\caption{\label{tabcbs}Fitting results for the cosmological models by using the CBS data. We quote $\pm 1\sigma$ errors, but for the parameters that cannot be well constrained, we quote the 95.4\% CL upper limits. Here, $H_0$ is in units of ${\rm km}/{\rm s}/{\rm Mpc}$ and $m_{\nu,{\rm{sterile}}}^{\rm{eff}}$ is in units of eV. Note also that the minimal value of $\chi^2$ for the $\Lambda$CDM model is $\chi^2_{\rm min}=3806.994$.}
\centering
\begin{tabular}{ccccccccccccccccccc}

\hline \multicolumn{1}{c}{} &&\multicolumn{3}{c}{No sterile neutrinos}&&\multicolumn{3}{c}{Massless sterile neutrinos}&&\multicolumn{3}{c}{Massive sterile neutrinos}&\\
           \cline{3-5}\cline{7-9}\cline{11-14}
Parameter  && $\Lambda$CDM & $w$CDM & HDE &&  $\Lambda$CDM & $w$CDM & HDE && $\Lambda$CDM & $w$CDM & HDE\\
\hline
$N_{\rm eff}$&&...&...&...&&$<3.46$&$<3.47$&$<3.58$&&$<3.45$&$<3.40$&$<3.56$\\
$m_{\nu,{\rm{sterile}}}^{\rm{eff}}$&&...&...&...&&...&...&...&&$<0.579$&$<0.651$&$<0.479$\\
$H_0$&&$67.69\pm0.43$&$68.31^{+0.83}_{-0.82}$&$67.86^{+0.79}_{-0.80}$&&$68.67^{+0.64}_{-1.02}$&$69.05^{+0.96}_{-1.13}$&$68.90^{+1.00}_{-1.20}$&&$68.15^{+0.50}_{-1.00}$&$68.78^{+0.91}_{-1.06}$&$68.64^{+0.96}_{-1.30}$\\
$w$&&...&$-1.029\pm0.033$&...&&...&$-1.020^{+0.034}_{-0.033}$&...&&...&$-1.037\pm0.036$&...\\
$c$&&...&...&$0.636^{+0.027}_{-0.029}$&&...&...&$0.649^{+0.028}_{-0.032}$&&...&...&$0.639^{+0.029}_{-0.033}$\\
$\sigma_8$&&$0.810\pm0.007$&$0.819\pm0.013$&$0.794\pm0.012$&&$0.818^{+0.008}_{-0.010}$&$0.824\pm0.014$&$0.800\pm0.013$&&$0.796^{+0.024}_{-0.014}$&$0.801^{+0.026}_{-0.018}$&$0.789^{+0.021}_{-0.016}$\\
\hline
$H_0$ tension && $4.27\sigma$ & $3.48\sigma$ & $3.80\sigma$ &&  $3.44\sigma$ & $2.91\sigma$ & $2.95\sigma$ && $3.91\sigma$ & $3.11\sigma$ & $3.14\sigma$\\
$\Delta\chi^2$  && 0 & $-0.534$ & 19.162 && $-0.240$  & $-0.585$ & 18.909 && $-0.272$ & $-0.617$ & 19.249\\
$\Delta \rm AIC$  && 0 & 1.466 & 21.162 && 1.760  & 3.415 & 22.909 && 3.728 & 5.383 & 25.249\\
\hline
\end{tabular}
\end{table*}

\begin{table*}\small
\setlength\tabcolsep{0.02pt}
\renewcommand{\arraystretch}{1.2}
\caption{\label{tabcbsh}Fitting results for the cosmological models by using the CBSH data. We quote $\pm 1\sigma$ errors, but for the parameters that cannot be well constrained, we quote the 95.4\% CL upper limits. Here, $H_0$ is in units of ${\rm km}/{\rm s}/{\rm Mpc}$ and $m_{\nu,{\rm{sterile}}}^{\rm{eff}}$ is in units of eV. Note also that the minimal value of $\chi^2$ for the $\Lambda$CDM model is $\chi^2_{\rm min}=3824.875$.}
\centering
\begin{tabular}{ccccccccccccccccccc}

\hline \multicolumn{1}{c}{} &&\multicolumn{3}{c}{No sterile neutrinos}&&\multicolumn{3}{c}{Massless sterile neutrinos}&&\multicolumn{3}{c}{Massive sterile neutrinos}&\\
           \cline{3-5}\cline{7-9}\cline{11-14}
Parameter  && $\Lambda$CDM & $w$CDM & HDE &&  $\Lambda$CDM & $w$CDM & HDE && $\Lambda$CDM & $w$CDM & HDE\\
\hline
$N_{\rm eff}$&&...&...&...&&$3.51\pm0.17$&$3.38^{+0.16}_{-0.20}$&$3.53\pm0.19$&&$3.54\pm0.18$&$3.35^{+0.12}_{-0.28}$&$3.55\pm0.20$\\
$m_{\nu,{\rm{sterile}}}^{\rm{eff}}$&&...&...&...&&...&...&...&&$<0.121$&$<0.413$&$<0.089$\\
$H_0$&&$68.25\pm0.42$&$69.77^{+0.75}_{-0.74}$&$69.37\pm0.71$&&$70.64^{+0.99}_{-0.98}$&$70.92\pm0.93$&$71.09^{+0.98}_{-0.99}$&&$70.59^{+1.01}_{-1.00}$&$70.80^{+0.95}_{-1.07}$&$71.10\pm1.00$\\
$w$&&...&$-1.078\pm0.031$&...&&...&$-1.047\pm0.033$&...&&...&$-1.068\pm0.038$&...\\
$c$&&...&...&$0.600\pm0.024$&&...&...&$0.636^{+0.027}_{-0.032}$&&...&...&$0.631^{+0.029}_{-0.032}$\\
$\sigma_8$&&$0.807\pm0.007$&$0.833^{+0.012}_{-0.013}$&$0.807\pm0.012$&&$0.829^{+0.010}_{-0.011}$&$0.839\pm0.013$&$0.816\pm0.013$&&$0.821^{+0.015}_{-0.012}$&$0.822^{+0.025}_{-0.016}$&$0.812\pm0.014$\\
\hline
$\Delta N_{\rm eff}>0$ && ... & ... & ... &&  $2.73\sigma$ & $1.67\sigma$ & $2.55\sigma$ && $2.74\sigma$ & $1.09\sigma$ & $2.52\sigma$\\
$H_0$ tension&& $3.90\sigma$ & $2.65\sigma$ & $2.94\sigma$ &&  $1.96\sigma$ & $1.83\sigma$ & $1.70\sigma$ && $1.97\sigma$ & $1.89\sigma$ & $1.69\sigma$\\
$\Delta\chi^2$  && 0 & $-6.659$ & 15.148 && $-6.471$  & $-8.904$ & 11.190 && $-6.497$ & $-9.089$ & 10.736\\
$\Delta \rm AIC$  && 0 & $-4.659$ & 17.148 && $-4.471$  & $-4.904$ & 15.190 && $-2.497$ & $-3.089$ & 16.736\\

\hline
\end{tabular}
\end{table*}

\subsection{The case of massless sterile neutrinos}

In the universe, the total energy density of radiation is given by
$$\rho_r=[1+N_{\rm eff}\frac{7}{8}(\frac{4}{11})^{\frac{4}{3}}]\rho_\gamma,$$
where $\rho_\gamma$ is the photon energy density. The effective number of relativistic species in the standard three-generation neutrino cosmology is $N_{\rm eff}=3.044$~\cite{Akita:2020szl,Froustey:2020mcq,Bennett:2020zkv}.

The presence of sterile neutrinos in the universe leads to $\Delta N_{\rm eff}=N_{\rm eff}-3.044>0$. In the case of massless sterile neutrinos, the only parameter of sterile neutrinos is $N_{\rm eff}$ and so we take the {fit result} of $\Delta N_{\rm eff}>0$ as the {preference of the existence} of massless sterile neutrinos.

In the Planck 2018 results \cite{Aghanim:2018eyx}, the case of taking $N_{\rm eff}$ as a free parameter is considered, and for the $\Lambda$CDM+$N_{\rm eff}$ model the result of $N_{\rm eff}=2.92^{+0.36}_{-0.37}$ (95\% C.L., Planck TT,TE,EE+lowE) is obtained. Obviously, if we set the requirement of $N_{\rm eff}>3.044$ for this case, then only a 95\% C.L. upper limit for $\Delta N_{\rm eff}$ can be obtained. Therefore, {the existence of massless sterile neutrinos is not preferred by} using only the Planck 2018 temperature and polarization power spectra.

Here, in the aspect of observational data, we consider BAO, SN, and $H_0$ data in addition to the Planck 2018 CMB data, and in the aspect of cosmological models, we consider $w$CDM and HDE in addition to $\Lambda$CDM. We wish to see what results can be given.

In Table~\ref{tabcbs}, we show the fitting results given by using the CBS data. We can clearly see that in this case for all the models only an upper limit on $N_{\rm eff}$ can be obtained, indicating that {the existence of massless neutrinos is not favored} by using the CBS data, no matter what dark energy model is considered.


Changes happen when the $H_0$ local measurement is added in the data combination. Table~\ref{tabcbsh} shows the case of using the CBSH data. We find that in this case {$\Delta N_{\rm eff}>0$ is favored at $2.73\sigma$, $1.67\sigma$, and $2.55\sigma$ level for $\Lambda$CDM+$N_{\rm eff}$, $w$CDM+$N_{\rm eff}$, and HDE+$N_{\rm eff}$}, respectively. 

Thus, we find that {the existence of massless sterile neutrinos is preferred} when the $H_0$ local measurement is included in the data combination. This is because $N_{\rm eff}$ is in positive correlation with $H_0$ when using the CMB data. As a result, a higher $H_0$ prior leads to a result of higher $N_{\rm eff}$. {The correlation between $N_{\rm eff}$ and $H_0$ can be clearly seen in Fig.~\ref{figneff}.}

In addition, the results of using AIC as an assess tool to compare dark energy are also shown in the tables. We can clearly see that the HDE model is definitely excluded by the current observations, since its $\Delta$AIC values are too high (around 23 in the case of CBS and around 15 in the case of CBSH). This confirms the previous results in Refs.~\cite{Guo:2018ans,Xu:2016grp}. We find that the $w$CDM model can well fit the current data.
In particular, we notice that, in the case of CBSH, $\Delta {\rm AIC}=-4.9$ for $w$CDM+$N_{\rm eff}$. {We also find that $w<-1$ is favored at around 1$\sigma$ level ($w=-1$ is outside the 1$\sigma$ limit), indicating that the $\Lambda$CDM model is not preferred by the CBS and CBSH data.}


\subsection{The case of massive sterile neutrinos}

In the case of massive sterile neutrinos, two extra parameters, $N_{\rm eff}$ and $m_{\nu,{\rm{sterile}}}^{\rm{eff}}$, need to be considered in the cosmological models. Here, the requirement of $N_{\rm eff}>3.044$ still holds.

From Table~\ref{tabcbs}, we can see that when using the CBS data neither $N_{\rm eff}$ nor $m_{\nu,{\rm{sterile}}}^{\rm{eff}}$ can be determined, no matter what dark energy model is considered. Only upper limits on $N_{\rm eff}$ and $m_{\nu,{\rm{sterile}}}^{\rm{eff}}$ can be obtained.

From Table~\ref{tabcbsh}, we can see that, when the $H_0$ measurement is included in the data combination, for all the dark energy models, $N_{\rm eff}$ can be effectively constrained. But even in this case, $m_{\nu,{\rm{sterile}}}^{\rm{eff}}$ still cannot be well constrained, and only upper limits can be given.

The {preference} of $\Delta N_{\rm eff}>0$ is at $2.74\sigma$, $1.09\sigma$, and $2.52\sigma$ level for the $\Lambda$CDM+$N_{\rm eff}$+$m_{\nu,{\rm{sterile}}}^{\rm{eff}}$ model, the $w$CDM+$N_{\rm eff}$+$m_{\nu,{\rm{sterile}}}^{\rm{eff}}$ model, and the HDE+$N_{\rm eff}$+$m_{\nu,{\rm{sterile}}}^{\rm{eff}}$ model, respectively, in the case of using the CBSH data.

For the constraints on $m_{\nu,{\rm{sterile}}}^{\rm{eff}}$, we find that the impact of dark energy is somewhat evident. Compared with $\Lambda$CDM, the constraint in $w$CDM is evidently looser, and that in HDE is slightly tighter. This also confirms the previous results in e.g. Refs.~\cite{Zhang:2015uhk,Feng:2017mfs,Zhang:2020mox}. We also show the main {results} in Fig.~\ref{figms}.

As the same in the above subsection, we also find that the HDE model is excluded by the current data because its AIC values are very high in the cosmological fits. The $w$CDM model is favored by the data,
{and $w=-1$ is also outside the $1\sigma$ limit, indicating that the $\Lambda$CDM model is not preferred by the CBS and CBSH data in the case of massive sterile neutrinos.}

\subsection{The Hubble tension}

We wish to check if the Hubble tension can be effectively relieved when sterile neutrinos are considered.

From Tables~\ref{tabcbs} and \ref{tabcbsh}, we can see that in the $\Lambda$CDM model (without considering sterile neutrinos) the $H_0$ tension is at 4.3$\sigma$ level for the case of using the CBS data, and at 3.9$\sigma$ level for the case of using the CBSH data. Thus, considering the $H_0$ measurement as a prior can slightly relieve the Hubble tension, but the tension still exists for $\Lambda$CDM at around 4$\sigma$ level.

In the case of using the CBSH data, we find that the HDE model with considering sterile neutrinos is the best one in relieving the Hubble tension (the tension is relieved to about 1.7$\sigma$). However, since the HDE model is excluded by the observational data (as assessed by the AIC tool), we do not consider this model in this issue. Actually, for the $\Lambda$CDM and $w$CDM models with sterile neutrinos, we find that they can also relieve the Hubble tension to less than 2$\sigma$. We also find that such models are favored by the current data. However, although the $H_0$ tension can be relieved to some extent, the $\sigma_8$ tension is slightly worsened (from 0.81 to about 0.83--0.84 in the case of massless sterile neutrinos; massive sterile neutrinos can slightly relieve this to about 0.82).

The Hubble tension is by no means easy to be resolved (see e.g. Refs.~\cite{Feng:2019jqa,Zhang:2019cww,Ding:2019mmw,Liu:2019awo,Guo:2019dui,Guo:2018ans,Gao:2021xnk,Yao:2020pji,Vagnozzi:2019ezj,DiValentino:2019ffd,DiValentino:2021izs,Freedman:2021ahq,Perivolaropoulos:2021jda} for various endeavors). When we remove the $H_0$ local measurement from the data combination, i.e., using the CBS data, we find that the Hubble tension is still at 3--4$\sigma$ level even sterile neutrinos are considered in the cosmological models.

\section{Conclusion}\label{sec4}

We use the Planck 2018 CMB anisotropy data combined with the latest BAO, SN, and $H_0$ data to search for sterile neutrinos in cosmology. In addition to the $\Lambda$CDM model, we also consider the $w$CDM model and the HDE model to show how properties of dark energy affect the {fit results} of sterile neutrinos.

We use the data combinations of CBS and CBSH to constrain the cosmological models. Based on the {constraints}, we can get the following conclusions.

(i) In the case of using the CBS data, no {preference of existence} of sterile neutrinos can be given, because only the upper limits on $N_{\rm eff}$ can be obtained. 
In this case, the $w$CDM model is most favored by the data; in the $w$CDM model, {$w=-1$ is outside the $1\sigma$ limit, indicating that the $\Lambda$CDM model is not preferred by the data}; the HDE model is definitely excluded by the current data because its AIC values are very high.

(ii) In the case of using the CBSH data, $\Delta N_{\rm eff}>0$ is {favored} at about 2.7$\sigma$ level for the $\Lambda$CDM model and at about 1--1.7$\sigma$ level for the $w$CDM model. But even in this case, $m_{\nu,{\rm{sterile}}}^{\rm{eff}}$ still cannot be well constrained and only upper limits can be given. 


(iii) When the $H_0$ local measurement is included in the data combination, the cosmological models of considering sterile neutrinos can effectively relieve the Hubble tension, with the tension relieved to less than 2$\sigma$. However, this does not mean that the Hubble tension can truly be resolved by this consideration, because when the local measurement of $H_0$ is removed in the data combination, the $H_0$ tension is then restored to about 3--4$\sigma$.

\begin{acknowledgments}
This work was supported by the National Natural Science Foundation of China (Grant Nos. 11947022, 12103038, 11975072, 11835009, 11875102, and 11690021),
the Liaoning Revitalization Talents Program (Grant No. XLYC1905011),
the Fundamental Research Funds for the Central Universities (Grant No. N2005030),
the National Program for Support of Top-Notch Young Professionals,
(Grant No. W02070050),
the Natural Science Foundation of Liaoning Province (Grant Nos. 2021-BS-154 and 2021-BS-156),
the 2019 Annual Scientific Research Funding Project of the Education Department of Liaoning Province (Grant No. LJC201915),
the Doctoral Research Project of Shenyang Normal University (Grant No. BS201844),
and the Natural Science Foundation of Shaanxi Provincial Department of Education (Grant No. 20JK0683).

\end{acknowledgments}

\end{document}